\documentclass[preprints,article,accept,moreauthors,pdftex,10pt,a4paper]{mdpi}
\newcommand{\ket}[1]{\left\vert#1\right\rangle}
\newcommand{\bra}[1]{\left\langle#1\right\vert}

\firstpage{1} 
\pubvolume{xx}
\issuenum{1}
\articlenumber{1}
\pubyear{2018}
\copyrightyear{2018}
\history{Received: date; Accepted: date; Published: date}

\Title{Coexistence of different scaling laws for the entanglement entropy in a periodically driven system}

\Author{Tony J. G. Apollaro $^{1,\dagger}$\orcidA{} and Salvatore Lorenzo $^{2,\dagger}$\orcidB{}}

\AuthorNames{Tony J. G. Apollaro and Salvatore Lorenzo}

\address{%
$^{1}$ \quad Department of Physics, University of Malta, Msida MSD 2080, Malta; tony.apollaro@um.edu.mt\\
$^{2}$ \quad Dipartimento di Fisica e Chimica, Universit\'a  degli Studi di Palermo, via Archirafi 36, I-90123 Palermo, Italy}

\corres{Correspondence: tony.apollaro@um.edu.mt; Tel.: +356-2340-8101}

\abstract{The out-of-equilibrium dynamics of many body systems has recently received  a burst of interest, also due to experimental implementations. The dynamics of both observables, such as magnetization and susceptibilities, and quantum information related quantities, such as concurrence and entanglement entropy, have been investigated under different protocols bringing the system out of equilibrium. In this paper we focus on the entanglement entropy dynamics under a sinusoidal drive of the tranverse magnetic field in the 1D quantum Ising model. We find that the area and the volume law of the entanglement entropy coexist under periodic drive for an initial non-critical ground state. Furthermore, starting from a critical ground state, the entanglement entropy exhibits finite size scaling even under such a periodic drive. This critical-like behaviour of the out-of-equilibrium driven state can persist for arbitrarily long time, provided that the entanglement entropy is evaluated on increasingly subsytem sizes, whereas for smaller sizes a volume law holds. Finally, we give an interpretation of the simultaneous occurrence of  critical and non-critical behaviour in terms of the propagation of Floquet quasi-particles.}


\begin{document}
\section{Introduction}
Entanglement in many-body systems is ubitiquos and both its characterisation and quantification is of paramount importance for many applications, ranging from quantum information processing tasks to the range of validity of several numerical methods based on tensor network algorithms, such as DMRG, MPS, and PEPS~\cite{RevModPhys.82.277}.
Amongst all possible types of entanglement, that between bipartite systems of a global pure state is one of the most investigated, due to the existence of a unique, well-established measure: the von Neumann entropy~\cite{RevModPhys.80.517}, dubbed, in such a context, {\textit{entanglement entropy}} (EE). This has yield to the investigation of the EE properties particularly in the ground state of many-body spin models~\cite{1751-8121-42-50-504002}. It has been shown~\cite{1742-5468-2007-08-P08024} that the EE of 1D gapped quantum systems described by a local Hamiltonian follows an area law, whereas, at criticality, a $\log$ law holds~\cite{PhysRevLett.90.227902}. On the other hand the EE obeys a volume law for general scrambled pure quantum states, for a class of pure states called cTPQ states~\cite{Fujita:2017pju}, and for typical excited states of quadratic Hamiltonians, including the quantum Ising model in 1D~\cite{Vidmar:2018rqk}. 
Recently, the out-of-equilibrium dynamics of many-body systems has received considerable attention~\cite{RevModPhys.89.011004}. As a consequence, also the dynamics of the EE of a quantum many-body system brought out-of-equilibrium by means of a quench~\cite{DeChiara:2005wb, PhysRevA.75.052321, 10.21468/SciPostPhys.4.3.017} or a periodic drive has been throughout investigated~\cite{PhysRevB.94.134304, PhysRevB.94.214301,1742-5468-2016-7-073101}. A universal behaviour of the EE, i.e., independent of the details of the protocol pushing the system out of equilibrium, has been observed. This includes a transient linear growth in time of the EE, settling in the asymptotic steady state into a volume law, meaning a linear scaling of the EE with the subsystem size. 

In this article we show that these different scaling laws can coexist in a periodically driven system depending on the size of the block we are considering. In other words, starting from a gapped (gapless) ground state, which exhibits an area (log) law, the EE of the periodically driven state, after a finite number of cycles, still obeys an area (log) law for large block sizes, whereas for small block sizes the volume law is attained. The crossover between the two scaling laws depends on the number of cycles and is derived by exploiting the quasi-particles propagation pictured proposed in Ref.\cite{1742-5468-2005-04-P04010}.
\section{Results}
We consider the 1D quantum Ising model, driven by a periodic
transverse magnetic field,
\begin{equation}\label{E.XYHam}
\hat{H}(t)=-\sum_{i=1}^{N}\left(\hat{\sigma}_i^x\hat{\sigma}_{i+1}^x-h(t)\hat{\sigma}_i^z\right)~,
\end{equation}
where $\hat{\sigma}^{\alpha}$ ($\alpha=x,y,z$) are the Pauli
matrices and $h(t){=}h{+}\Delta h \sin\left(\omega t\right)$ is the
harmonically modulated  transverse field. This model
belongs to the Ising universality class and it exhibits a  second order
quantum phase transition with the critical point located at $h{=}h_c{=}1$,  separating a ferromagnetic phase from a
paramagnetic  one.
The model depicted in Eq.~\ref{E.XYHam} with a static field is diagonalised by standard
Jordan-Wigner and Bogolyubov transformations~\cite{LIEB1961407}, and results in a free fermion spinless Hamiltonian $\hat{H}=\sum_k
\varepsilon_k\left(2\hat{\gamma}_k^{\dagger}\hat{\gamma}_k-1\right)$.
Here $\varepsilon_k{=}\sqrt{\left(h-\cos k\right)^2+\left(\sin k\right)^2}$ is the energy of the Bogolyubov quasiparticle
with momentum $k$, and annihilation operator
$\hat{\gamma}_k{=}u_k\hat{c}_k{+}v_k\hat{c}_{-k}^{\dagger}$, the
$\hat{c}_k$'s being JW fermion annihilation operators labelled by the
momentum $k$.
The ground state of the Hamiltonian can be finally written in the BCS form $\ket{GS}\equiv\ket{\Psi(t{=}0)}{=}\prod_{k>0}\ket{\psi(0)}_k{=}\prod_{k>0}\left(u_k{+}v_k \hat{c}_k^{\dagger}\hat{c}_{-k}^{\dagger}\right)\ket{0}$, with $\ket{0}$ representing the vacuum of the JW fermions
($\hat{c}_k \ket{0} = 0 \, , \forall k$).

In order to determine the time-evolved state we make use of the Floquet formalism~\cite{GRIFONI1998229} which allows, in time-periodic Hamiltonians, to map the stroboscopic time evolution to a dynamics generated by a time-independent Hamiltonian, dubbed Floquet Hamiltonian, $\hat{H}_F$, and defined by $\hat{U}(T)=e^{-i T \hat{H}_F}$. For the model here under scrutiny see, e.g., Ref.~\cite{lorenzo_1} for the derivation of the unitary evolution operator $\hat{U}(T)$. Furthermore, exploiting translational invariance, we are able to write the state at times $t=nT$, where $n$ is the number of cycles, as $\ket{\Psi(nT)}=\prod_{k>0} \left ( u_k(nT) + v_k(nT)
\hat{c}^{\dag}_k \hat{c}^{\dag}_{-k} \right ) \ket 0 $.

Because of the gaussian nature of the time-evolved state, we just need to evaluate the correlation matrix $\Gamma$ for the block of $l$ spins in order to determine its EE~\cite{PhysRevA.75.052321}, where
\begin{equation}\label{E.CorrMat}
\Gamma(nT)=
\begin{vmatrix}
\alpha(nT) & \beta^{\dagger}(nT)\\
\beta(nT) & \mathbf{1}-\alpha(nT)
\end{vmatrix}~,
\end{equation}
with $\alpha(nT)$ and $\beta(nT)$ representing $l\times l$ matrices whose elements reads, $\alpha_{nm}(nT)=\bra{\Psi(nT)}\hat{c}_n \hat{c}^{\dagger}_m\ket{\Psi(nT)}$ and $\beta_{nm}(nT)=\bra{\Psi(nT)}\hat{c}_n \hat{c}_m\ket{\Psi(nT)}$, respectivly. From the correlation matrix $\Gamma$ the EE is readily evaluated: $S(l;nT)=-Tr\left[\Gamma(nT)\log_2 \Gamma(nT)\right].$ We refer the reader to Ref.~\cite{PhysRevB.94.134304} for the dynamical behaviour of $S(l;nT)$, whereas in the following section we show that different scaling laws for the EE coexist at finite times.
\section{Discussion}
\begin{figure}[ht!]
\includegraphics[width=1.\textwidth]{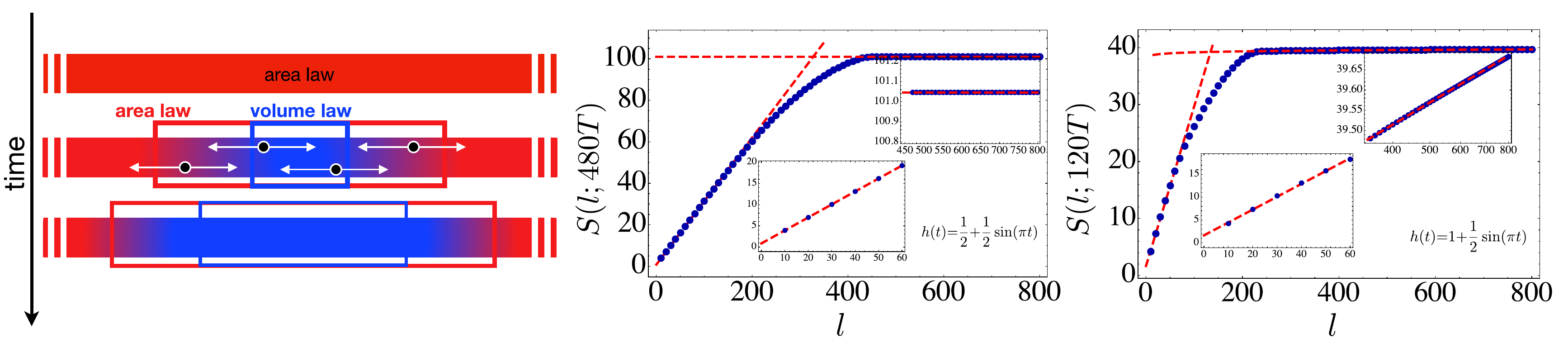}
\caption{(left) For a non-critical (critical) system at $t=0$ an area (log) law for the entanglement entropy is satisfied. At $t>0$, the periodic drive is responsible for the appearence of quasi-particle travelling along the chain entangling the spins in the region they have spanned, i.e., $l=2 v_{max}t$, and represented in the left figure by arrows. If the subsystem is chosen such that the arrows span over its whole size, every spin in it is entangled with spins outside and a volume law follows (blu box), otherwise an area (log) law holds (red box). A snapshot of the entanglement entropy for different subsystem sizes $l$ in a spin chain of $N=8192$ for a periodic drive far from criticality after $n=480$ cycles (left) and at criticality after $n=120$ cycles (right). The insets show the coexistence of a volume with an area and with a log law, respectively. } \label{F.qb5}
\end{figure}

\begin{figure}[ht!]
\begin{center}
\includegraphics[width=0.9\linewidth]{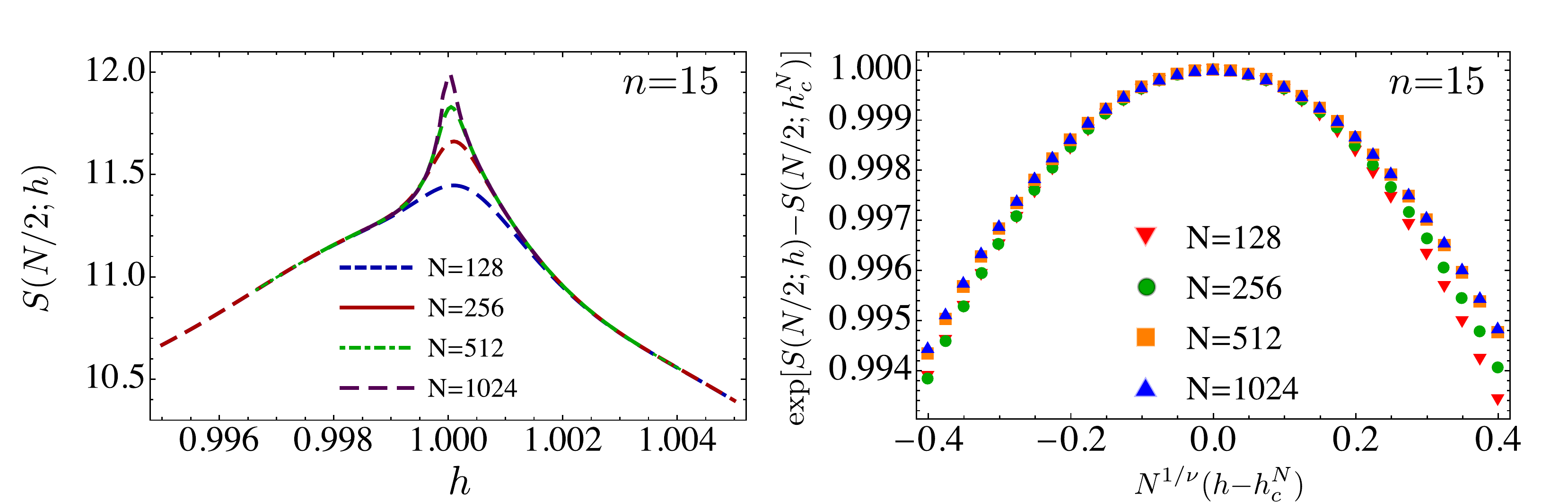}
\caption{(left) Logarithmic divergence with the system size $N$ of the half-chain entanglement entropy after $n=15$ cycles of the drive $h(t)=1+\frac{1}{2} \sin\left(\pi t\right)$. (right) Finite-size scaling with the critical exponent $\nu=1$ of the Ising universality class of the data on the left panel close to $h=h_c$.} \label{F.EEfsc}
\end{center}
\end{figure}

In the middle (right) panel of Fig.~\ref{F.qb5} we show the EE of different subsystem block sizes after $480$ ($120$) cycles, where the drive is applied to the ground state of a gapped (gapless) Hamiltonian. The insets show that the EE increases linearly with the size, yielding thus a volume law for small sizes, whereas an area (log) law is maintained for larger sizes.
A simple explanation of the coexistence of the area ($\log$) law and the volume law can be given in terms of quasiparticles produced at each site by the periodic drive, left panel of Fig.~\ref{F.qb5}. According to Ref.~\cite{1742-5468-2005-04-P04010} quasi-particles of opposite momentum $\{k,-k\}$ travelling along the chain with $v=\frac{d\mu_k}{dk}$, where $\mu_k$ are the eigenenergies of $\hat{H}_F$, are responsable for the linear growth of the entanglement enropy. After a time $t{=}l/2v_{max}$, all spins residing in the block $l$ are entangled with spins residing outside the block $l$, as a consequence a volume law appears for $l_1 < l$. On the other hand, for $l_2\gg l$ only the spins in the region $\tilde{l}$ at the boundaries  are entangled with spins belonging to the complementary region, and hence an area (log) law follows. In between $l_1$ and $l_2$ a transitory region where neither a volume nor a area (log) law applies.
Note that, notwihstanding the increase of the EE under periodic drive~\cite{PhysRevB.94.134304}, the latter occurs in such a way to fulfill the log law and, more importantly, it obeys the finite size scaling Ansatz~\cite{1742-5468-2008-06-P06004} typical of critical systems at equilibrium $S_{\frac{N}{2}}(h){-}S_{\frac{N}{2}}(h_c^N){=}f\left(N^{\frac{1}{\nu}}\left(h-h_c^N\right)\right)$, Fig.~\ref{F.EEfsc}. A similar scenario does appear also for local boundary drivings~\cite{PhysRevLett.118.260602}. However, this simple picture does not hold for other quantities, such as nearest-neighbor concurrence and susceptibility, where the time scale of the persistence of critical-like behaviour under periodic drive requires a different explanation in terms of the fidelity of low-energy modes, see Ref.~\cite{lorenzo_1}. The results here obtained may be experimentally verified in periodically driven ultracold atoms in optical lattices~\cite{RevModPhys.89.011004}.

\acknowledgments{ T.J.G.A. and S.L. acknowledge support from the Collaborative Project
QuProCS (Grant Agreement 641277)}
\authorcontributions{T.J.G.A. and S.L. contributed equally to the paper.}
\conflictsofinterest{The authors declare no conflict of interest. The founding sponsors had no role in the design of the study; in the collection, analyses, or interpretation of data; in the writing of the manuscript, and in the decision to publish the results.} 
\reftitle{References}
\externalbibliography{yes}
\bibliography{biblio.bib}
\end{document}